# Knowledge Integration of Collaborative Product Design Using Cloud Computing Infrastructure

Mahdi Bohlouli[#1], Alexander Holland[#2], Madjid Fathi[#3]

[#]*Department of Science and Engineering, University of Siegen*
*Hoelderlinstr. 3, D-57076 Siegen, Germany*

[1]mbohlouli@informatik.uni-siegen.de
[2]alex@informatik.uni-siegen.de
[3]fathi@informatik.uni-siegen.de

*Abstract*— **The pivotal key for the success of manufacturing enterprises is sustainable and innovative product design and development. In collaborative design, stakeholders are heterogeneously distributed chain-like. Due to the growing volume of data and knowledge, an effective management of the knowledge acquired in the product design and development is one of the key challenges facing most manufacturing enterprises. Opportunities for improving efficiency and performance of IT-based product design applications through centralization of resources such as knowledge and computation have increased in the last few years with maturation of technologies such as SOA, virtualization, grid computing, and/or cloud computing. The main focus of this paper is the concept of ongoing research in providing the knowledge integration service for collaborative product design and development using cloud computing infrastructure. Potentials of the cloud computing to support the Knowledge integration functionalities as a Service by providing functionalities such as knowledge mapping, merging, searching, and transferring in product design procedure are described in this paper. Proposed knowledge integration services support users by giving real-time access to knowledge resources. The framework has the advantage of availability, efficiency, cost reduction, less time to result, and scalability.**

## I. INTRODUCTION

Changes made during the early design stage do not cause the significant increase in costs, while during the production stage, sharp increase in costs will occur since many blueprints, design documents or components would require re-work and re-design [5]. Today's research is focused on optimising the development methodologies to enable shorter time, lower costs and higher quality of the systems [2]. The pivotal key for the success of manufacturing enterprises is sustainable and innovative product design and development. In order to achieve this goal, it is required to have a real and deep knowledge of former and current procedures in the manufacturing enterprise [4] and future needs as well as customer feedbacks and various stages of production chain activities. Realization of an efficient knowledge transfer between different stakeholders of product development process such as linking customers and suppliers proactively throughout the entire value chain, and collaborating across boundaries in distributed enterprises is reinforcing this trend. Due to the growing volume of data, information and knowledge, an effective management of the knowledge gained in the product development is one of the key challenges facing most enterprises [5]. In addition, organizations are more successful that are considering not only using the acquired knowledge but also innovating based on knowledge and managing knowledge-driven ideas.

Opportunities for improving efficiency and performance of IT-based product design applications through centralization of computation and resources such as knowledge and information using maturation of technologies like SOA, virtualization, grid computing, and cloud computing have been increased in the last few years.

Cloud Computing [6, 9] is an elastic execution environment of resources involving multiple stakeholders. It provides a scalable, reliable, and highly cost effective service at multiple granularities for a specified level of quality (of service) [8]. Enterprises can create new services by dynamically provisioning of compute, storage, and application services and offer as their own isolated or composite cloud services to users [9]. Depending on the use case, security, and financial issues: services can be deployed in public, private or hybrid clouds. Enterprises may use cloud functionalities from others for their internal computing or to offer their own services to users outside of the company based on provisioned resources of public clouds. In private clouds whereby enterprises set up cloud-like, centralized shared infrastructure with automated capacity adjustment that internal departmental costumers utilize in a self-service manner. Hybrid clouds [8] consist of a mixed employment of a public and private cloud to achieve a maximum of cost reduction through outsourcing whilst maintaining the desired degree of control over e.g. sensitive data by employing local private clouds.

A collaborative product design [10] environments have two common capabilities: distribution and collaboration. Distribution means that system consists of geographically dispersed members who are working to fulfil a global design objective. The act of working together means collaboration. Applying new IT technologies such as Web 2.0, cloud-based solutions and so on would increase the efficiency and effectiveness of collaboration in product design. In collaborative design, centralization of resources such as enterprise global knowledge plays fundamental role to get higher efficiency and smarter collaboration.

Knowledge integration (KI) [5] consists of identification, acquisition, evaluation and utilization of external knowledge. It is different from Knowledge Management (KM). Identification is about knowing the type of knowledge which is needed. If user knows what knowledge he needs, the way to get it is acquisition process of KI. Knowledge utilization means making the acquired external knowledge in the form of usable for internal aims. KI consists of tools and techniques to help the system to identify, acquire, and utilize external and internal knowledge.

The main focus of this paper is the concept of ongoing research in providing the knowledge integration service for collaborative product design and development using cloud computing infrastructure. The potential of cloud computing to support the Knowledge Integration as a Service (KIaaS) by providing functionalities such as knowledge mapping, merging and some other knowledge services such as search, and transfer in product design procedure are studied in this paper. Product design knowledge is firstly raw and scattered data located in different distributed parts of the design chain. It could be explicit or tacit. Related works and research projects are studied at the next section. Product knowledge sources, collaborative design process and classification of product knowledge are then described.

## II. RELATED WORKS

Y.J. Chen and Y.M. Chen [11] proposed a distributed product knowledge service model based on product lifecycle and its supply chain. The proposed model is distributed, modular, flexible and product-oriented. It includes three different layers: Physical Knowledge Layer, Business Logic Layer and User Layer. Business Logic layer itself covers Product Knowledge Service Functional Module, System Management & Control Functional Management Module and Collaborative Organization Formation Functional Module. These modules afford services such as product knowledge query, representation, reasoning, navigating, acquiring, merging, mapping, managing and so on. It can be applied to other knowledge-intensive systems such as medical diagnostics.

Y. Jiang et al [1] proposed an ontology-based framework of knowledge integration to support business process in collaborative manufacturing. They defined the ontology for product knowledge, which is composed of product design, process planning and manufacturing knowledge. These stages include some related sub-concepts; ontology schema is then designed. Similarity matching method has been applied in order to achieve ontology integration. Concept name, essential information and relationship similarity calculation is used in the mapping process. The framework supposed to effectively integrate an individual enterprise's knowledge by providing comprehensive concepts and knowledge connections, and increase reuse ratio of product knowledge and reduce product development cost and cycle time.

E. M. Kern and W. Kersten [13] introduced a framework for internet supported inter-organizational product development. The factor which has been highlighted in this study is efficiency. The efficiency and effectiveness of their framework is based on choosing the appropriate level of partner integration. They used internet as a basis for collaboration between partners.

E. Revilla and T. Curry [14] developed a framework to test the impact of customer and supplier knowledge integration in product development performance. They highlighted the positive effects of customer in product innovation. As a result, trust and learning culture influence knowledge integration capability in product development. Survey methodology is used in this research. Product development performance is then measured through two distinct components: Teamwork values process outcomes and value to customer expresses product outcomes.

Jian and colleagues [17] developed a knowledge search mechanism using autonomous, intelligent agents to transform passive search and retrieval engines into active, personal assistants of multi-agent product development systems. That aims to facilitate knowledge retrieval in order to prompt right knowledge for design and analysis. Improving the performance of short-term information retrieval in existing search or retrieval engines is the result of this project.

Toussaint et al [15] proposed a methodology to capture manufacturing knowledge and its application towards the design verification and validation of new engineering designs. They outlined that capturing proper knowledge methodology could decrease the time of new product development and increase its quality. The approach is implemented into web-based PLM prototype and Computer Aided Design (CAD) system. The PLM system chosen for implementing the proposed approach is a web-based PLM prototype called ACSP. This PLM system allows experts to transcribe their knowledge and users to operate independently throughout the methodology.

Mahesh et al [16] studied a collaborative tool for virtual manufacturing. They provided the framework to relay Computer Aided Manufacturing (CAM) information and knowledge between all manufacturing chains. The internal structure of each functional agent is modularized into several components to ensure the compatibility and upgradeability of the system. In this project, different functional agents of geographically distributed and collaborative manufacturing agents can interact coherently. With specific agents having unique functionalities, a Manufacturing Managing Agent (MMA) acts as the centre of this distributed manufacturing system.

Consideration of scalability in IT supported services of product design and development process is important due to the growing volume of product knowledge and information. In addition, existing knowledge of manufacturing enterprise should always be available and accessible to the users of system all over the world through different types of platforms such as computers, PDAs, manufacturing machineries and so on. None of above studied projects considered scalability, availability and accessibility. The use of cloud computing could benefit these factors. As a result, manufacturing enterprises will be able to support peak times processing without any need to buy further hardware or software solutions. They can provision (rent) resources provided in the cloud and release them after finishing the usage in order to have the advantage of scalability. Since cloud services support different platforms through virtualization and other techniques, the use of cloud as a service sources could also benefit accessibility of KI service.

## III. PRODUCT KNOWLEDGE AND DESIGN PROCESS

To study the product knowledge and related processes, it is required to have an understanding of what knowledge is. In practice, the terms knowledge, information and data are often used interchangeably, but giving a correct definition of knowledge is possible by distinguishing among knowledge, information and data [20]. A commonly held view including minor variants is that data is raw numbers and facts, information is processed data [20, 21], and knowledge is the result of learning and reasoning. In fact Data is unclassified and unprocessed values, a static set of transactional elements such as 211102345, or structured records of transactions. Information is meaningful context-related collection of data which has been processed and organized. As an example knowledge is a set of understandings and the state of knowing acquired through experience or study to use in decision making activities [22] which includes facts, opinions, ideas, theories, principles, models, ignorance, awareness, familiarity, understanding, and so on. Knowledge can be explicit and/or tacit. Explicit knowledge is a type of knowledge that can be articulated, codified or stored, and/or represented in the media. Tacit knowledge emphasizes understanding the kinds of knowledge that individuals in an organization have. Information Systems (IS) can convert data into information by making them meaningful and context-related, but they are not capable of converting information to knowledge. Gathering, storing, providing effective search, retrieval mechanisms for locating relevant information, transferring knowledge, and to provide link among sources of knowledge are supported with IT in Knowledge Based Systems (KBSs) [19]. KBSs are tools to make intelligent and rule-based decisions based on an inference engine with justifications to support human decision-making, learning and action.

In collaborative design environments, engineers and designers can share their experience and knowledge with globally distributed colleagues via provided services. They can closely work with customers, manufacturers and other related users to get for instance their requirements and feedback to apply directly into design chain. This is very important to apply changes in the design step rather than manufacturing. It is possible to get active or passive feedback as well as customer and market needs for future products and other data and knowledge using provided services to optimise next generation products design and development. Figure 1 shows general view of collaborative product design and development over internet.

Operating principles, normal configurations, judgment skills, ways of doing and thinking, general product-specific design competence, and best practices are instances of captured knowledge in product design procedure. Table 1 shows the most related knowledge resources to product design and development.

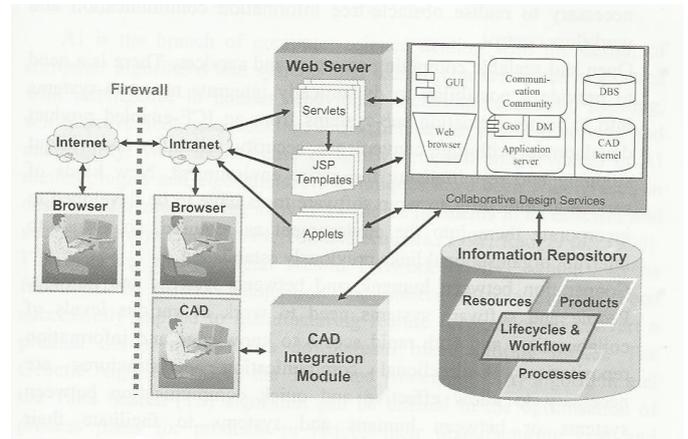

Fig. 1 Collaborative Product Design and Development [10]

In collaborative product design and development, knowledge resources can be internal or external. Internal ones are located inside of an enterprise. Employees and personal archives are a type of internal knowledge. External knowledge resources are located outside of an enterprise such as the knowledge came from partners or acquired through cloud services. In proposed framework knowledge could be local or global. Local knowledge is the knowledge stored inside of an agent. Knowledge which is located outside of an agent, but has been acquired through cloud services and stored in local servers is considered as a local knowledge. Global knowledge is the knowledge stored on the cloud and shared to agents and development phases as a public service.

TABLE 1

SOURCES OF NEW PRODUCT DESIGN AND DEVELOPMENT KNOWLEDGE [5]

| Rank | Source |
|------|--------|
| 1 | Customer (most important) |
| 2 | Specialized Magazines |
| 3 | Production Employees |
| 4 | Staff |
| 5 | Suppliers |
| 6 | Sellers |
| 7 | Brochures and catalogues |
| 8 | Industrial fairs |
| 9 | Commercial fairs |
| 10 | Business fairs (less important) |

Product Knowledge includes all activities of product life cycle. These activities are not only production procedure activities, but also pre and post production activities such as idea generation, product design, customer feedback, support, and recycling activities. Reuse of acquired knowledge and expertise will cut re-experimenting costs and time, but optimum knowledge reuse needs the proper selection of the resources in right time and place for appropriate aim and target. Considering the large mass of activities, product knowledge increases gradually along with manufacturing procedure. Figure 2 shows the general classification of product knowledge into 3 subgroups: Product Specifications Knowledge (PSK), Product Technical Knowledge (PTK), and Product Processes Knowledge (PPK).

### A. Product Specifications Knowledge (PSK)

Product Specifications Knowledge (PSK) includes general identifications of the product such as name, type, owner information, product items, and history. This group of knowledge is defined with production manager and product planner. The knowledge captured in this group is a type of explicit knowledge. Information such as series of product generations and first generation date and identification of the previous and next generations of the product are known as PSK.

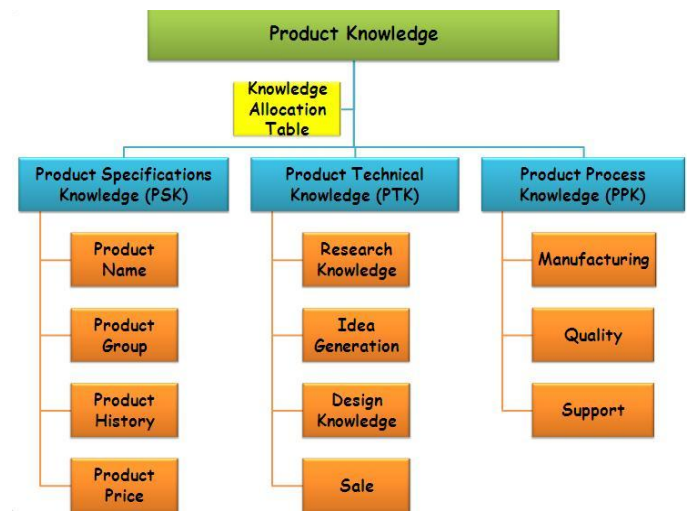

Fig. 2 Product Knowledge Classification

### B. Product Technical Knowledge (PTK)

Product Technical Knowledge (PTK) is the most knowledge intensive and useful part in product development. Knowledge acquired from different sources such as research phase, design process, idea generation and product sale and market analysis are PTK sources. Each phase and step has different types and large volume of knowledge. For instance, research institute consists of the knowledge about product testing procedures, computer simulation results, research instrumentalities, and specialized research competences.

Idea generation stage itself consists of close collaboration between partners and product planners. Knowledge-based innovation in this stage affects the whole production and the success of an enterprise. It is about new product ideas using active feedbacks, requirement analysis, market monitoring, and so on. The knowledge sources of this step could be internal or external. Companies with an external orientation alongside internal bearings are more likely to innovate. Marketing and sale department should monitor market developments to support creative idea generation. Customers, suppliers, distributors, and competitors are instances for external knowledge resources to be used in idea generation phase. Customers can benefit both product improvement and future product developments. Trends, needs and demands of market segments, new product/market combinations, cultural factors, and policy changes are some other instances of knowledge sources as PTK specially in idea generation phase. General information of product users, territorial demands and interests of products, available products and services, market demands and distribution activities are some prevalent resources of product knowledge in sale phase. PTK can be explicit or tacit. Product design knowledge which has been described earlier distinguished as PTK.

### C. Product Process Knowledge (PPK)

Product Process Knowledge (PPK) is acquired knowledge from manufacturing, quality control, failure analysis, and support activities. Properties of natural and artificial materials, ability to manufacture, logistics, capacity, production process and related activities are some instances of PPK during manufacturing. Competence in pilot production/ scale up, production scheduling and timing, and equipment layout are some other sources of manufacturing knowledge.

Quality Control and performance indicators are effective product knowledge resources in production system to improve the quality of current and future products. Performance of components or materials in pilot, production quality control, failures reported during the production and their analysis, preventive activities details, condition monitoring data and gathered data from different monitoring sensors during the production are all quality related knowledge sources. In case that manufacturing processes and devices are being monitored using different types of sensors such as cameras, thermometers, ultrasonic detectors, temperature and pressure sensors, the gathered data and information from detectors and their analysis is important in product process KI. In addition, corrective activities to prevent machinery downtimes and detailed documentation of such experiences could benefit knowledge reuse in future preventive activities. Reported failures via diagnostic systems and relevant solutions, history and log files of all corrective activities are some other knowledge sources of PPK which could benefit current and future product and service development procedures.

Customer support and service activities result useful knowledge as PPK. As mentioned earlier they are used in different product development phases such as idea generation and so on. Feedback from customer, faced difficulties during support procedure, made service tickets, guarantee and warrantee information, time duration of reported failures and relevant solutions and all other information are support and service related knowledge sources. This step is in close connection to design and development process, while the change and reports of this process can make dramatic optimizations in next generation products. This type of knowledge is mostly explicit.

As mentioned in table 1 the most important knowledge about products is related to customers. [5] Manufacturing enterprises use mainly the knowledge of their close partners such as customers, employees, and suppliers. In collaborative design and development (especially in cloud supported environments) enterprises can acquire knowledge from partners. It is very important to reuse partner knowledge and information to benefit the whole production process.

## IV. CLOUD-BASED DESIGN KNOWLEDGE INTEGRATION

Knowledge and data integration of product design and development is data intensive processing; such applications need reliable and scalable services. To handle the problem, enterprises need to have powerful servers, storage and computing resources. They need to be aware of peak times processing and provide appropriate resources, but so far these resources will mostly be idle and will be used just in peak time processing. They have to pay for some additional costs such as installation, maintenance and administration overhead, cooling and power, upgrade, software licenses and software update costs and etc. Also it is time consuming to plan a suitable hardware and software infrastructure and to provide appropriate solutions. Any assignment/project signed up by the enterprise can kick-start quickly as the biggest head-ache of due-diligence in procuring database, hardware, software license are all taken care of by the cloud computing provider.

Provided real-time KI services are preconfigured, ready to use and accessible all over the world. As mentioned earlier, services have the advantage of scalability. Users can demand resources for peak time processing, this is significantly cost and time-effective; they could release resources after finishing the processing. In addition they could provide their own and customized services to third-party users based on provisioned services. As a result SMEs will have the advantage of using services with lower costs and shorter time to result. There is the possibility of knowledge sharing between different enterprises based on provided and standardized services. This will benefit knowledge transfer of inter and intra enterprises. Proposed framework supports high availability and dependability which are necessary engineering features for global always-on, always-available systems.

The architecture of proposed framework in this paper exists four layers: Cloud Presentation Layer (CPL), Cloud Knowledge integration Service Layer or Knowledge integration as a Service (KaaS), Cloud Physical Layer (CPhL) and Security Layer (SL). (Fig. 3) Each agent of the collaborative design enterprise has a GateKeeper (GK) which is the type of server(s) to recieve requests from local users and provide proper solutions. The solution could be based on local knowledge resources stored in GK or provisioned resources of cloud. The knowledge base of the GK is updated periodically and replicated into the cloud. GKs have local knowledge bases of the most used knowledge for local agent. There is a firewall between GKs and CPL because of security issues. GK is at the top of the architecture, which is set of computer hardware and software that is required to connect to the layers of the cloud. GK ideally includes the users of the cloud computing services.

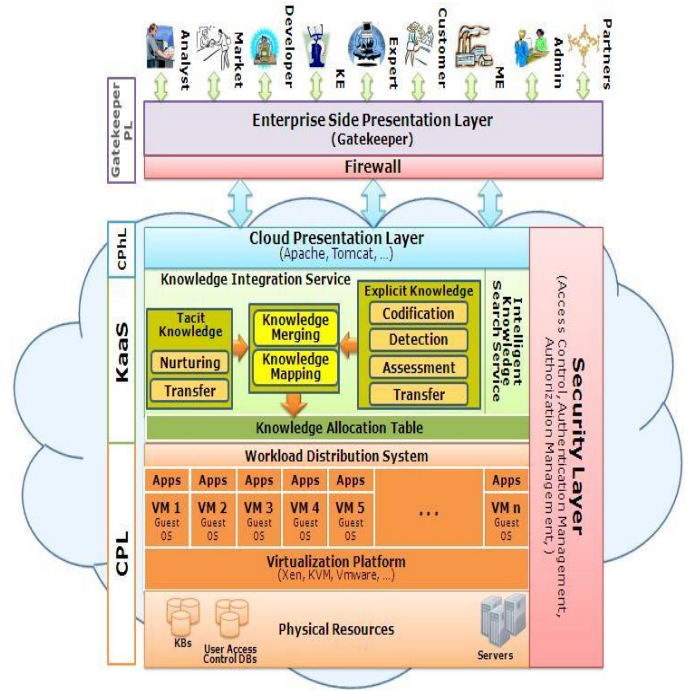

Fig. 3 Cloud Based Product Design Knowledge Integration Architecture

### A. Cloud Presentation Layer

Cloud Presentation Layer (CPL) is the most visible layer to the end-users of the system. Users access the services provided by this layer through web-portals. Manufacturing Engineers (MEs), customers, developers and manufacturers can reach assigned software and hardware resources via this layer for developing and deploying their applications on the cloud. CPL is secure and used for comminucating lower layers and outside world. This layer provides access to Knwoledge Engineers (KEs) to refresh, validate and verificate knowledge bases and do some other knowledge based processing. Design engineers (DEs) can access to online informations of ongoing design activities and relevant knowledge to be informed about the state of the design. This includes a full application, customization options, meeting enterprises requirements, etc.

### B. Cloud Knowledge Integration Service Layer

Cloud knowledge integration service layer or KKnowledge integration as a Service Layer(KaaS) provides the knowledge based services to the users of the sytem such as customers, manufacturers, KEs, MEs and developers. KaaS is the part that all KEs apply their knowledge based activities and process data. This layer is an interface between users (via CPL) and cloud physical layer including knowledge bases and servers. Users can request to get an access to DataWare Houses (DWH) and Knowledge Bases (KBs) and other deployed services and applications in the cloud after getting authenticated using cloud security layer. It is from this perspective that individuals, IT operations, experts and KEs deploy knowledge capabilities running on top of the CPL. KaaS includes knowledge integration and search services. Integration service of this layer supports knowledge mapping and merging. KI service for design process depends on the type of the knowledge (Tacit or Explicit). A. Jetter et al [5] described KI components clearly based on the type of knowledge: explicit knowledge integration includes codification, detection, assessment and transfer of knowledge. Codification [5] is articulation and transit of explicit knowledge from human source to any type of media such as texts, voices, videos. After codification, the knowledge is detached and could be transferred easily. This stage can be done automatically by KaaS. Detection process defines the location and importance of the knowledge.

We are supposed to give different values as an importance to the knowledge such as: Secure, High Priority, Normal, and Low Priority. Access control of users and knowledge requesters could be defined based on these importance values. Definition of such values will help us in providing knowledge sharing services. Detection step is important because it defines the source and location of the knowledge. This will help us in improving the functionality of knowledge search service. The most effective step of KI to the search process is assessment. Knowledge Assessment step is validating and weighing the knowledge. It is attaching credibility, value, significance, weight to the explicit knowledge. The weight values of the knowledge which is very important for search process includes three dimensions: Geographically Significance, Usability, popularity. All these values will be improved through usage of the system and knowledge.

Usability is the value which will be initialized by knowledge users. Each user whenever requests the knowledge and receives the proper solution, after the process of knowledge reuse, he will give the value of usability for given knowledge to his request in percents. Because of the different tasks of each stage in collaborative design, it is clear that the knowledge of one section may not be so useful for another section located somewhere else. In such situations geographical significance value has the key role in assigning the proper knowledge. As an example, when we have two types of knowledge with same or similar popularity and usability values, the knowledge with closer (higher) geographical significance to the requested place is more related to the request. The last step of product design explicit knowledge integration is transfer and includes two steps: first in the local enterprise resources and then replication of the knowledge and data into the cloud for security, backup and sharing issues.

Integration of tacit knowledge is difficult. It can only be captured when it is found. The key of successful integration of such knowledge is to accurately find the right expert or knowledge source to solve the specific problem. Enterprises can, by automatically capturing interactions such as communication channels, emails, instant messaging and so on, expand the scope of reusable knowledge to include data like the following:

- What content is helpful
- what processes have worked best for
- which type of issues
- what experts were involvedwhat pitfalls to avoid

This tacit knowledge is automatically captured, stored in the cloud knowledge base and immediately usable. So, the next time there is a similar critical situation in manufacturing enterprise, knowledge workers can use knowledge search service of the proposed method to find the exact source and can tap into time-saving, relevant information to increase the quality of resolution while reducing resolution time. Because partners are also able to deploy knowledge into the collaborated knowledge base, the selection range of knowledge source in this method is wider than traditional ones.

To integrate product design knowledge and effectively reuse it in future, first product knowledge ontology has been made based on section 3. This ontology is 2 levels: local ontology and global ontology. Local ontology is stored knowledge in GK of each stakeholder in distributed enterprise. This is because sometime required knowledge can be found at local sources and it is not required to use cloud services. Each GK whenever receives the request for knowledge first checks the local ontology. Global ontology is the one which has been deployed inside the cloud; it includes the local ontology of all connected gatekeepers and other partners.

Global ontology is much more complete than locals, it is cloud based service. The next step of integration is knowledge mapping. The method of mapping is based on [1] with some modifications. Similarity calculation of knowledge essentials has been used. The most important task in this step is use of history and wider knowledge domain. The accuracy of matching method depends on the size of KB. Because in collaborative design and cloud based services, the source is centralized and is a collection of acquired knowledge from whole system and partners, search and selection domain is greater. As a result accuracy is higher than traditional one.

*C. Cloud Physical Layer*

Cloud Physical Layer (CPhL) is the base layer which works mainly for clouding. As its name suggests in this layer are all physical devices that run on the cloud, halls and corridors with thousands of computers stacked on top of each other. Some of the computers will be used to store data, others to process information, to store other files, etc. It runs state-of-the-art hardware that can be multiplexed by the virtual infrastructure level to make the physical virtual. CPhL has the basic computing resources and machines, operating systems, and storage. A current example of this layer is Amazon Elastic Compute Cloud (EC2). Knowledge Bases, DBs, Application servers, mainframes, grids, clusters, storages, are included in this layer. CPhL may use virtualization technologies such as Xen, KVM, VMWare to scale more and support multi-platform possibility.

Data of the different cloud users, which would be available in various geographical locations, is backed up on a regular basis as a preventive mechanism against crisis situations. Cloud computing can further be enhanced to deliver the virtual middleware level as a service on top of the Physical layer. A virtual machine that can be spun up within a couple of minutes is a key to the ability to provide for the demand.

Virtualization also forces the removal of specialized equipment on which software may depend by providing a baseline, non-specialized, machine abstraction. Multi-tenancy is a further concept in this layer, where multiple customers can share a single physical resource at the same time. Virtual machines running on the same physical hardware are an example of multi-tenancy. In order to make all of these shared, virtualized resources available on demand, some automation tools need to set behance the request for a resource and the fulfilment of the request.

Databases in this layer are distributed, replicated, and largely transactional. These can be separated from the rest of the cloud stack through RESTful APIs between different vendors but there is a definite latency advantage to coupling of data and its interpreter.

### D. Cloud Security layer

Any of the cloud layers can be subjected to security attacks. To increase the degree of dependability and ensure that there is no loss, data and infrastructure are hosted on mirror sites and have a replicated copy at all times. Enterprise's data is stored in locations close to its competitor's data. The security layer takes care that all data stored in the cloud are encrypted and has limited access to authorized users. This layer has a risk management plan in place for each layer. To support information security, public key encryption, X.509 certificates, and the Secure Sockets Layer (SSL) is used to enable secure authentication and communication. For credential management, system stores and manages the credentials for a variety of systems and users can access them according to their needs. Secure and safe storage of credentials is equally important.

## V. CONCLUSIONS

We expect a growing importance for knowledge based distributed and collaborative product design using cloud computing. This means that enterprises are interested to use their design services with more flexibility and scalability. In addition to these factors reduction in costs and time can be achieved through the use of cloud based services. Knowledge based activities are data intensive and they need powerful resources. In this paper we studied general classification of product knowledge as three groups: Product Specification knowledge, Product Technical Knowledge and Product Process Knowledge. We discussed different types of product knowledge which is included in these groups such as operating principles, normal configurations, judgment skills, ways of doing and thinking, general and product-specific design competence, and best practices. We studied collaborative product design process and activities. General cloud based architecture is then proposed. This framework includes four different layers: Cloud Presentation Layer, Cloud Knowledge Integration Service Layer, Cloud Physical Layer, and Cloud Security Layer. Cloud Presentation layer is the gate which users can have an access to cloud services and lower layers. Knowledge service layer includes different components based on the type of knowledge and finally knowledge mapping and merging. Knowledge Integration layer also includes intelligent knowledge search functionality. Both services are able to read and write in the knowledge allocation table to help better knowledge search and classification.

Physical layer is the most important layer of cloud based framework, in fact it is the most cloud related layer. Enterprises may use virtualization techniques to benefit multi platform supports. Load balancing between different servers is also supposed in this layer. Data of the different cloud users, which would be available in various geographical locations, is backed up on a regular basis as a preventive mechanism against crisis situations. As a result, proposed framework points following advantages described earlier:

- Reduce Production Cost and Time
- Providing the meta-knowledge model
- Increase the level of customization
- Further integration of customer needs and feedback into manufacturing activities specially design process
- Availability of large number of quantities of design knowledge in real-time and from multiple sources
- Increased degree of dependability using replicated copy of data and knowledge all times
- Increased accessibility across many platforms including mobile
- Improved efficiency from high utilization of sharing physical servers
- Improved reliability with replication of data within the system and higher level of fault tolerant

In future, virtual objects such as manufacturing robots, sensors and machinery will be able to connect provisioned servers of cloud computing directly without any need to human interactions and engineering activities. In fact future objects will connect together using internet computing and there will not be any distance between geographically distributed factories.

### ACKNOWLEDGMENT

We would like to thank Shelley Smith and Amir Tabatabae for their constructive comments on an earlier version of this paper.

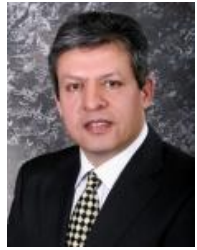

Madjid Fathi received the Ph.D. degree in Mechanical Engineering from the University of Dortmund in Germany. He is a full Professor of Computer Science at the University of Siegen. His research interests are in the areas of Computational Intelligence, Knowledge Based Systems, Applications in Medicine and Mechanical Engineering, Applied Knowledge Management.

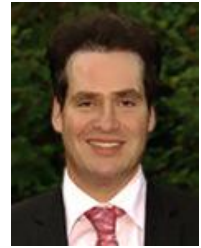

**Alexander Holland** received the Ph.D. degree in Computer Science from the University of Siegen in Germany. He is a Post Doctoral researcher at the University of Siegen. His research interests are in the areas of Computational Intelligence, Graphical Modelling, Decision Making and Decision Support Systems, Distributed Systems, Applied Soft Computing, Relationship Discovery in Knowledge Management

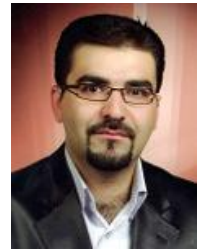

Mahdi Bohlouli received the Masters degree in Computer Systems Architecture from the Iran University of Science and Technology at Tehran, Iran.
He is a research assistant and PhD candidate at the University of Siegen. His research interests are in the areas of Knowledge Integration and Modelling, Cloud and Parallel Computing and Distributed Information Systems**.**